\documentstyle[12pt]{article} 
\input epsf
\newcommand{\be}{\begin{equation}} 
\newcommand{\ee}{\end{equation}} 
\newcommand{\bea}{\begin{eqnarray}} 
\newcommand{\eea}{\end{eqnarray}}

\begin{document} 
 
\vspace{0.5in} 
\begin{center} 
{\large\bf The Wilson Effective K\"ahler Potential For Supersymmetric  
Nonlinear Sigma Models} 
\end{center} 
\begin{center} 
{\bf T.E. Clark and S.T. Love}\\ 
{\it Department of Physics\\  
Purdue University\\ 
West Lafayette, IN 47907-1396} 
\end{center} 
\vspace{1.0in} 
\begin{center} 
{\bf Abstract} 
\end{center} 
 
Renormalization group methods are used to determine the  
evolution of the low energy Wilson effective 
action for supersymmetric nonlinear sigma  
models in four dimensions.  For the case of supersymmetric $CP^{N-1}$ models, 
the  K\"ahler potential is determined exactly and is shown to exhibit 
a non-trivial ultraviolet fixed point in addition to a 
trivial infrared fixed point.  The strong 
coupling behavior of the theory suggests the possible  
existence of additional relevant operators or nonperturbative
degrees of freedom. 
\pagebreak

The action, $\Gamma$, for four dimensional supersymmetric nonlinear 
sigma models is determined by the K\"ahler potential, 
$K(\phi,  \bar\phi)$ \cite{Z}, and takes the form 
\be 
\Gamma =\int dV K(\phi, \bar\phi). 
\label{action} 
\ee 
Such models naturally emerge in the analysis of many 
underlying supersymmetric gauge theories when describing the physics 
below a scale $\Lambda$  
where the gauge nonsinglet degrees of freedom are confined but yet above the 
supersymmetry breaking scale. In that case, 
($\bar\phi^{\bar i}$)  $\phi^i$ denote the 
relevant light gauge singlet (anti-) chiral superfield 
degrees of freedom. Alternatively, 
($\bar\phi^{\bar i}$)  $\phi^i$ could be the Nambu-Goldstone superfields 
resulting from the spontaneous breakdown of an internal symmetry group $G$ 
to an unbroken subgroup $H$ at a scale 
$\Lambda$, at which the supersymmetry is still unbroken \cite{BLPY}. 
In either case, the effective action containing all terms through 
two space-time derivatives is the supersymmetric nonlinear 
sigma model \cite{BW}-\cite{BL} action
of Eq.~(\ref{action}). 
 
The self radiative corrections to the action can be determined 
by functionally integrating over the degrees of freedom below the 
scale $\Lambda$ \cite{W}\cite{HW}. Thus for  supersymmetric 
nonlinear sigma models, the Wilson effective K\"ahler potential at scale 
$\Lambda(t)=e^{-t}\Lambda,~~ t>0$ is obtained 
by integrating out the degrees of freedom 
between $\Lambda$ and $e^{-t}\Lambda$. The resulting effective action at 
this lower scale can be equivalently used to describe the physics on all 
lower energy scales. While the non-renormalization theorem 
\cite{FL}\cite{GRS} guarantees the absence of induced 
superpotential terms, the effective 
action will, in general, contain corrections to the K\"ahler potential as well 
as terms containing higher powers of space-time derivatives. 
These higher derivative terms will be 
consistently neglected in the subsequent analysis. Although this is a 
truncation of the model, it is an improvement over the 
often used local approximation 
\cite{H}\cite{CHL} which completely neglects the radiative corrections to 
all terms containing derivatives. 
The reason we are able to go beyond the local 
approximation in the present case is a direct consequence of the supersymmetry 
which allows all terms containing up to two space-time derivatives 
to be derivable from a potential function \cite{CL2}. Thus our 
analysis allows a 
determination of the anomalous dimension for the (anti-) chiral superfield. 

Integrating out the  
degrees of freedom in an infinitesimal momentum shell just below the 
scale $e^{-t}\Lambda $ while rescaling all dimensionful parameters by 
$e^{-t}\Lambda $ and all fields according to their anomalous dimensionality,  
the compensating change in 
the K\"ahler potential can be characterized by a nonlinear partial 
differential equation in $t$ and the superfields which holds independent 
of the strength of the coupling.  
The solution to this Wilson (exact) renormalization group equation is 
tantamount to explicitly performing the functional integration into the 
infrared. 

For definiteness, we consider the action of Eq.~(\ref{action}) for the 
particular case of the 
supersymmetric $CP^{N-1}$ model. Here there are $N-1$ (anti-) 
chiral superfields, $(\bar{\phi}^{\bar{i}}) \phi^{i}~,~
i,\bar i =1,...,N-1 $, 
whose lowest components are the coordinates for the homogeneous 
K\"ahler manifold $\frac{SU(N)}{SU(N-1)\times U(1)}$. 
Defining the K\"ahler metric at scale $\Lambda(t)$ as 
\be
g_{i\bar i}\equiv K_{i\bar i}(\phi, \bar\phi) ~,
\ee
where the subscripts on $K$ 
denote differentiation with respect to the superfields, 
e.g. $K_{\bar i} ={\partial K\over \partial \bar\phi^{\bar i}}$, 
the Wilson renormalization group equation takes the form 
\bea 
{\partial g_{i\bar i}\over \partial t} &=& -2\gamma g_{i\bar i} -(1+
\gamma)\phi^j  g_{i\bar i j} -(1+\gamma)\bar\phi^{\bar j} g_{i\bar 
i\bar j} \cr 
 & & +{1\over 8\pi^2}\left[g_{i\bar i j\bar j} g^{-1}_{\bar j j}
-g_{i \bar j j}g_{k\bar i \bar k}g^{-1}_{\bar k j}g^{-1}_{\bar j k}\right],
\label{wrge1} 
\eea 
where $\gamma$ denotes the anomalous dimension for the $N-1$ chiral 
superfields. 
Since the fields are rescaled according to their anomalous dimensions 
as the system flows into the infrared, the chiral field anomalous 
dimension can be extracted by evaluating equation (\ref{wrge1}) at 
$\phi^i=0=\bar\phi^{\bar i}$ where $g_{i\bar i}\vert_{\phi^i 
=0=\bar\phi^{\bar i}}=\delta_{\bar{i} i}$ .  So doing, one finds that 
\be 
2\gamma \delta_{i\bar i}= {1\over 8\pi^2} 
\left[g_{i\bar i j\bar j} g^{-1}_{\bar j j}
-g_{i \bar j j}g_{k\bar i \bar k}g^{-1}_{\bar k j}g^{-1}_{\bar j k}\right]
\vert_{\phi^i =0=\bar\phi^{\bar i}} ~. 
\ee 
Assuming that the K\"ahler potential is a function of the product of 
chiral and antichiral superfields 
$\rho\equiv \bar\phi^{\bar i} \delta_{\bar i i}\phi^i$ with $\delta 
_{ii}=N-1$, the Wilson equation for the effective K\"ahler potential, 
$K(\rho,t)$, then reduces to 
\be 
{\partial K\over \partial t} = 2K -2(1+\gamma)\rho K_\rho +{1\over 
8\pi^2}\left[\ln{(K_\rho +\rho K_{\rho\rho})} +(N-2)\ln{K_\rho}\right] ~, 
\label{wrge2} 
\ee 
while the anomalous dimension of the chiral superfields is given by  
\be 
\gamma ={N\over 16\pi^2}K_{\rho\rho}\vert_{\rho=0}. 
\ee 
Note that the $\rho=0$ normalization of the metric translates into the 
normalization $K_\rho\vert_{\rho=0} = 1$. 
The partial differential equation (\ref{wrge2})
describes the exact renormalization 
group flow of the effective  
action for any pure chiral theory where the quantum radiative 
corrections have been truncated to include only the K\"ahler potential 
which are the terms that are at most of order $p^2$ in its momentum 
expansion.  

Contrary to most Wilson renormalization group equations which can only be 
treated by numerical means, Eq.~(\ref{wrge2}) admits the analytical solution 
\be
K(\rho,t)=\frac{1}{\chi(t)}\ln{(1+\chi(t)\rho)}
\label{KX}
\ee
where $\chi(t)$ is an effective coupling constant satisfying the 
renormalization group equation
\be 
{d\chi(t)\over dt} = -2(1+\gamma)\chi(t), 
\label{g} 
\ee 
and where the anomalous dimension is 
\be 
 \gamma (t)=-{N\over 16\pi^2 }\chi(t). 
\label{gamma} 
\ee
The form of the solution displayed  in Eq.~($\ref{KX}$) is not 
altogether unanticipated. For models with fields 
in homogeneous K\"ahler manifolds, the form of the K\"ahler potential is 
uniquely determined by the group structure of the $G/H$ coset space.
The Wilson renormalization group 
flow of the K\"ahler potential must then describe the exact 
renormalization group running of the finite number of coupling 
constants (decay constants) in such models.  For the case of the 
supersymmetric $CP^{N-1}$  model, there is one such coupling and 
the K\"ahler potential in terms of a particular set 
of coordinates can be written in the form given by Eq.~(\ref{KX}). 

Defining the scaled coupling constant $\kappa (t)$  as 
\be 
\kappa (t) \equiv {N\over 16\pi^2 }\chi(t) , 
\ee 
the renormalization group equation becomes 
\be 
{d\kappa\over dt}= -2\kappa (1-\kappa )  
\ee 
while the anomalous dimension is simply
\be
\gamma (t)=-\kappa (t)~.
\ee
Thus, within the higher derivative operator truncation 
we are working, the exact beta function for $\kappa$ 
takes the simple form 
\be
\beta_\kappa = 2\kappa (1-\kappa )
\ee
and is seen to be a sum of two terms only. We reiterate that in obtaining 
this result we have included contributions from an infinite number of 
operators. A similar form for the beta 
function has previously been advocated \cite{K} based on extrapolating the 
results of a $(2+\epsilon )$ expansion calculation\cite{P}\cite{BZ}. 
For two dimensional supersymmetric 
sigma models such a behavior for the beta function has also been previously 
argued \cite{AGF}\cite{NSVZ} using an alternate chain of reasoning 
and demonstrated explicitly by calculation 
through four-loop order \cite{GVZ}.

The beta function displays an ultraviolet fixed point at $\kappa =\kappa_c 
\equiv 1$, 
in addition to the trivial infrared fixed point. The renormalization 
group equation is readily integrated producing the explicit solution 
\be 
\kappa (t)= {\kappa (0) \over (1-\kappa(0))e^{2t}+\kappa (0)} .
\ee 
Alternatively expressed, the ultraviolet fixed point is indicative of the 
existence of a phase trasition with an associated renormalization 
group invariant inverse correlation 
length \cite{K} 
$\xi^{-1}=\Lambda (\frac{1}{\kappa(0)}-\frac{1}{\kappa_c})^{\frac{1}{2}}$ 
satisfying
\be 
\xi^{-1} =\Lambda(t)\left({1\over \kappa(t)}-{1\over 
\kappa_c}\right)^{\nu^{\prime}}, 
\label{corr} 
\ee 
where $\nu^\prime ={1\over 2}$ is a critical exponent.  

\begin{figure}[ht]
\epsfysize=240pt \epsfbox[-360 0 0 800]{zflow.ps}
\caption{ Renormalization group flow of the coupling constant $\kappa$.  The 
arrows denote the coupling constant evolution as the system flows into 
the infrared.  An ultraviolet non-trivial fixed point ocurrs at $\kappa 
= \kappa_c \equiv 1$.}
\label{fig:Fig1}
\end{figure}

Figure 1 illustrates the various renormalization group flows. 
For negative bare couplings, $\kappa(0)<0$, 
the theory evolves from a Landau singularity in 
the ultraviolet towards the trivial fixed point in the infrared. Note that 
these flows admit a simple particle interpretation with a free field theory  
emerging in the far IR. On the other hand, the flows for positive bare coupling 
either approach the trivial fixed point if the bare coupling 
is less than the UV fixed point, $\kappa(0)<\kappa_c=1$, or run away if 
$\kappa(0)>\kappa_c =1$. However, for each of these flows, the scalar 
superfield propagator goes as $\frac{\Lambda(t)^{-2\gamma}}{p^{2-2\gamma}}$  
with a negative anomalous dimension, 
$\gamma <0$. This signals a violation of unitarity.  
In order for this pathological behavior to be 
circumvented, the model must either contain additional relevant or marginal 
operators or admit additional nonperturbative degrees of freedom whose 
presence will reverse the sign of the 
anomalous dimension for positive bare coupling.  
The fact that higher derivative operators may 
not be irrelevant in the 
context of $2 + \epsilon $ expansions for sigma models has been 
discussed in \cite{We}\cite{CC}\cite{C} 
and could have some bearing on this issue. 
Indeed it is just this class of operators which we have neglected in our 
current analysis. Whether the relevance of these operators is simply an 
artifact of the $2 + \epsilon $ expansion or could be invalidated by 
higher order calculation is still an open question. 
In addition, it is known that in lower dimensions the $CP^{N-1}$ model 
asdmits nontrivial classical solutions; instantons in d=2 and
monopoles in d=3.  Thus one might speculate that the lack of a 
simple particle interpretation for positive bare coupling 
for the d=4 supersymmetric $CP^{N-1}$ model is 
reflective of the need to include string degrees of freedom.

\bigskip 
 
The authors wish to thank Serguei Khlebnikov for numerous enjoyable 
and useful discussions.  This work was supported in part by the U.S. 
Department  
of Energy under grant DE-FG02-91ER40681 (Task B). 
\pagebreak

\newpage 

\begin{thebibliography}{99}
\bibitem{CHL}T.E. Clark, B. Haeri and S.T. Love, 
{\it Nucl. Phys.} {\bf B402} (1993) 628. Note that the sign convention for 
the field anomalous dimension in this paper is opposite 
from that currently employed.
\end{thebibliography}
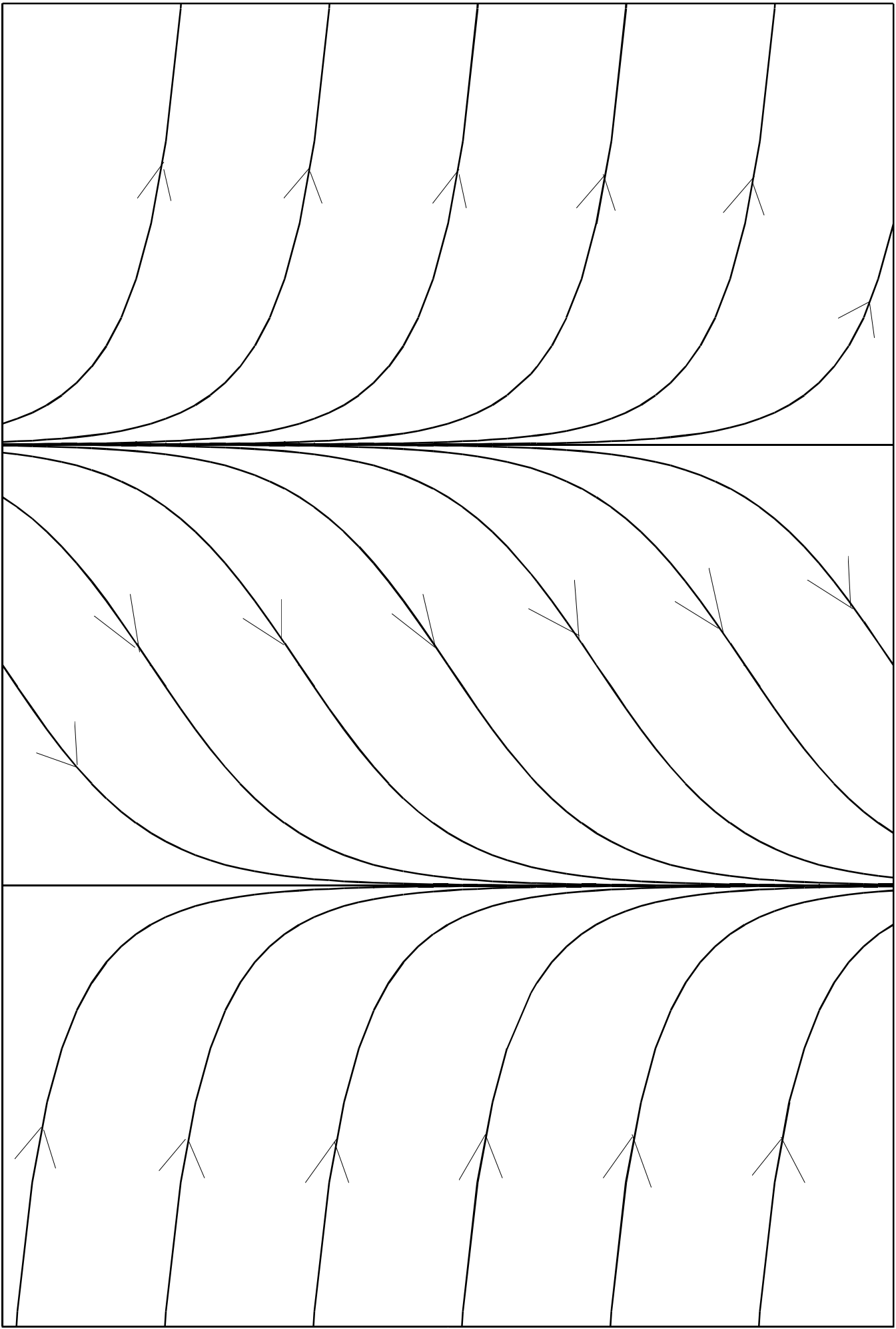
\end{document}